# Probing angle dependent thermal conductivity in twisted bilayer MoSe$_2$

Manab Mandal[1,2], Nikhilesh Maity[3], Prahalad Kanti Barman[2], Ashutosh Srivastava[3], Abhishek K. Singh[3], Pramoda K. Nayak[2,4*], Kanikrishnan Sethupathi[1,2*]

[1] Low temperature physics lab, Department of Physics, Indian Institute of Technology Madras, Chennai 600 036, India
[2] Department of Physics, Indian Institute of Technology Madras, Chennai 600 036, India
[3] Materials Research Centre, Indian Institute of Science, Bangalore-560012, India
[4] Centre for Nano and Material Sciences, Jain (Deemed-to-be University), Jain Global Campus, Kanakapura, Bangalore 562112, India

Twisted bilayer (t-BL) transition metal dichalcogenides (TMDCs) attracted considerable attention in recent years due to their distinctive electronic properties, which arise due to the moiré superlattices that lead to the emergence of flat bands and correlated electron phenomena. Also, these materials can exhibit interesting thermal properties, including a reduction in thermal conductivity. In this article, we report the thermal conductivity of monolayer (1L) and t-BL MoSe$_2$ at some specific twist angles around two symmetric stacking AB (0°) and AB′ (60°) and one intermediate angle (31°) using the optothermal Raman technique. The observed thermal conductivity values are found to be $13 \pm 1$, $23 \pm 3$, and $30 \pm 4$ W m$^{-1}$K$^{-1}$ for twist angle [θ] = 58°, 31°and, 3° respectively, which is well supported by our first-principles calculation results. The reduction in thermal conductivity in t-BL MoSe$_2$ compared to monolayer ($38 \pm 4$ W m$^{-1}$K$^{-1}$) can be explained by the occurrence of phonon scattering caused by the formation of a moiré superlattice. Herein, the emergence of multiple folded phonon branches and modification in the Brillouin zone caused by in-plane rotation are also accountable for the decrease in thermal conductivity observed in t-BL MoSe$_2$. The theoretical phonon lifetime study and electron localization function (ELF) analysis further reveals the origin of angle-dependent thermal conductivity in t-BL MoSe$_2$. This work paves the way towards tuning the angle-dependent thermal conductivity for any bilayer TMDCs system.

## I. INTRODUCTION

Fourier's law of heat conduction is based on the assumption of a continuous temperature field, which is valid only for macroscopic systems where the number of particles is very large. At the microscopic scale (individual atoms or molecules), the concept of temperature becomes more complex, and the continuous temperature field assumption breaks down because of the dimensional limitation [1], thermal rectification [2], and ballistic transport [3]. Hence, the study of thermal conductivity in two-dimensional (2D) limits is significant for both fundamental science [1–3] as well as in technological applications [4–6]. On the other hand, 2D materials, including graphene and TMDCs, are frequently used in optoelectronic [4], biological monitoring [5], and energy storage [6] devices because of their outstanding optical, electrical, and mechanical characteristics. These materials also possess unique thermal properties that exhibit notable distinction compared to their bulk counterparts due to the reduced dimensionality and the confinement of phonons [7]. In order to enhance heat dissipation in optoelectronic devices, it is essential to study the thermal conductivities of these materials. Recently Balandin *et al*. reported lower thermal conductivity in monolayer TMDCs compared to a few layers one, which corresponds to the inverse trend seen in the widely researched 2D materials, for example, in graphene [8]. Therefore, in order to understand this ambiguity in thermal conductivity values of 2D materials with reference to layer number, more rigorous experiments and theoretical models are required. Particularly in their few-layer versions [9–23], TMDCs are extremely fascinating for use in electrical, optical, and thermal applications. Despite extensive research on electron transport in well-explored TMDCs materials like $MoS_2$, $WS_2$, $WSe_2$, and $MoSe_2$, the number of experimental reports [24] published on thermal transport is very limited. In addition, $MoSe_2$ exhibits low thermal conductivity compared to other TMDC materials, which makes it an excellent thermoelectric material [25]. Compared to monolayer $MoS_2$ and $WS_2$, the value of thermal conductivities of monolayer $MoSe_2$ predicted from atomistic simulations show inconsistent values, varying from 17.6 [25] to 54 W m$^{-1}$K$^{-1}$ [26]. Most of the investigations reported so far concentrate on either monolayer, bilayer or bulk $MoSe_2$. However, a systematic variation in thermal conductivity with reference to twist angle is missing in the literature. Moreover, the thermal conductivity of a system with a twisted bilayer (t-BL) is affected by the rotation of the atomic planes to the increased number of phonons scattering sites, which is due to the incorporation of moiré superlattice. The change in the phonon dispersion due to moiré superlattices leads to the emergence of phonon minibands, which can be used to enhance phonon transport in specific directions. In this study, we report the thermal conductivities of atmospheric pressure chemical vapour deposition (APCVD) grown monolayer (1L) $MoSe_2$ as well as t-BL $MoSe_2$ using the optothermal Raman technique [8,27] for selected twist angles. The observed thermal conductivity values are found to be 13 ± 1, 23 ± 3 and, 30 ± 4 W m$^{-1}$K$^{-1}$ for twist angle [θ] = 58°, 31°and, 3° respectively. In the case of 1L, the thermal conductivity has been revealed to be 38 ± 4 W m$^{-1}$K$^{-1}$, which is larger than the bilayer. Such variation in thermal conductivity was approximated using first-principles calculations, which supports our experimental findings. This study paves the way to explore the twist angle-dependent thermal conductivity for other bilayer TMDCs as well as heterostructures (HSs).

## II.  MATERIAL AND METHODS

Atmospheric pressure chemical vapour deposition (APCVD) was used to grow 1L $MoSe_2$ flakes, as reported in our earlier study [28]. In a typical run, a ceramic boat containing $MoO_3$(60 mg) precursors was positioned in a heating zone centre of a quartz tube furnace of 2 inches in diameter. In the same quartz tube, a quartz boat was positioned at upstream (in the low-temperature zone) and filled with 100 mg of Se powder. Two pieces of sapphire substrates were positioned at downstream with a gap of 2 mm and were placed on the ceramic boat containing $MoO_3$ powder. A typical temperature of 700 °C and 285 °C were used to heat the $MoO_3$ and Se precursors, respectively, with an increased rate of 25 °C per minute. The growth period was set for 45 minutes, and the Ar: $H_2$ was kept at 60:12 sccm during the growth process. Finally, in the presence of an Ar atmosphere, the furnace had been brought down to ambient temperature.

The technique of aided transfer using Poly (methyl methacrylate) (PMMA) was employed to transfer the 1L $MoSe_2$ flakes over $SiO_2$ (300 nm) /Si substrate (1×1 cm$^2$), which have been grown in their natural state on sapphire. The $MoSe_2$/sapphire substrate was subsequently spin-coated for 60 seconds at 4000 rpm, after which a few drops of PMMA (677 °C) were applied until the entire surface was covered. The sample had been spin-coated and then put on a heated plate for roughly 30 minutes at 120 °C, and around all four edges of the substrate with PMMA coating have been scraped with a sharp blade up to 1 mm inside to provide an easy-to-clean border and simple passage for gripping the substrate from all the sides. In order to facilitate the separation of PMMA from the sapphire substrate, a tiny amount of DI water was applied to the PMMA /sapphire interface. The hydrophobic characteristic of $MoSe_2$ prevents the water drop from covering the surface; instead, it assists in hydrating at the interface of sapphire and PMMA-coated $MoSe_2$ flakes. Using a pointed tweezer, PMMA-coated $MoSe_2$ film was pilled from the sapphire substrate. For better cleanliness, it was rinsed twice with DI water for two minutes each time. Then the floating PMMA/$MoSe_2$ was put onto the $SiO_2$/Si substrate. The PMMA coating was subsequently put away by submerging the sample in warm acetone for an entire hour. Here, the required $MoSe_2$ on $SiO_2$ (300 nm)/Si was eventually achieved. In order to get t-BL $MoSe_2$, the above process was repeated with the use of as prepared $MoSe_2$/($SiO2$/Si) as substrate instead of ($SiO_2$/Si) followed by cleaning of PMMA layer using warm acetone. In this process, bilayer MoSe$_2$ flakes with different stacking orientations were fabricated.

Optical microscopy (OM) images were captured using a Nikon (ECLIPSE LV100ND) microscope with a 100X objective lens and a NA of 0.95. Room temperature low-frequency Raman (LFR) measurements were performed using a 'confocal Raman spectrometer' (Renishaw, in Via Reflex). The spectrometer has a 'backscattering geometry' under the same 532 nm laser excitation. Using an ultra-low frequency filter and a 100X objective with a numerical aperture (NA) of around 0.8, LFR spectra have been taken with 1800 lines/mm grating. The high-frequency Raman spectra have been measured at ambient and high temperatures making use of a confocal Raman spectrometer (LabRam 800, Horiba Jobin Yvon) having backscattering geometry. The measurements were carried out with a 50X (NA ~ 0.55) with a long working distance along with a 633 nm laser excitation. The laser spot diameter of around 1.4 μm was employed to concentrate on the sample, and the spectrum was collected by using 1800 lines/mm grating. The samples have been placed in a liquid nitrogen-cooled Linkam cryostat for temperature dependent measurement. The laser power is kept below 14 μW during the entire temperature dependent measurements.

The first-principles DFT computations have been completed by applying the Vienna ab initio simulation package (VASP) [29,30] for monolayer and different stacking orders of bilayer MoSe$_2$. For the purpose of describing the ion-electron interactions within the system, all-electron projector augmented wave (PAW) potentials [31,32] are implemented. The electronic exchange and correlation component of the potential can be illustrated through the Perdew-Burke-Ernzerhof (PBE) [33] generalized gradient approximation (GGA). To avoid the spurious interactions between periodically repeated images and make the 2D sheet of monolayer and bilayer, a 15 Å vacuum along the c-axis is used. In the basis sets for planar waves, the Kohn-Sham (KS) orbitals have been expanded with an energy cut-off of 500 eV. The lattice parameters and atomic positions are optimized by taking into account the weak vdW interactions between the layers, which are included in Grimme's PBE-D2 [34], where the pair-wise force field characterizes the weak vdW interactions. For relaxation, the Brillouin zone of the structures is sampled using a well-converged, Γ-centered Monkhorst-Pack (MP) [35] k-grid of $12 \times 12 \times 1$ for the monolayer and bilayers of MoSe$_2$. The conjugate gradient approximation is used to optimize the monolayer and bilayers until the Hellmann-Feynman forces operating on each individual atom are smaller than 0.005 eV/Å. Applying the principles of density functional perturbations (DFP) [36], phonon dispersion is determined. Phonopy [37], an extra package with a VASP interface that is based on the Parlinski-Li-Kawazoe approach, has been utilized for phonon dispersion. The Lattice thermal conductivities ($\kappa_{latt}$) can be obtained by solving the Boltzmann transport equation (BTE) for the semi-classical phonon, where the anharmonic interatomic force constants (IFCs) are calculated by considering all the interactions up to the third nearest neighbours using Phono3py package [38]. The harmonic second and third-order interatomic forces are computed with high accuracy of energy cut-off 500 eV and energy convergence criterion of $10^{-8}$ eV. The super cells of $3 \times 3 \times 1$ are considered to calculate the second and third-order forces. The calculations of force constants are obtained by sampling the BZ with a $5 \times 5 \times 1$ k-grid. For $\kappa_{latt}$, a converge q-grid of $19 \times 19 \times 1$ is used for monolayer and bilayer MoSe$_2$. The Born-effective charges are taken into account for the long-range interactions, and the DFP principle is used to compute them.

### III. EXPERIMENTAL RESULTS AND DISCUSSION

The current literature raises a number of questions about the heat transmission process in layered MoSe$_2$ as well as other TMDCs [6,39]. The thermal conductivity of these systems exhibits significant changes during the transition from 1L to BL as well as multi-layer systems. Although thermal conductivity is well studied in BL TMDCs, it is limited for the twisted configuration in the case of TMDCs. Recent study in twisted bilayer black phosphorus (BP) shows a significant twist-induced change in thermal conductivity [40]. The anisotropy involved in the case of BP might be the cause of non-uniform thermal conductivity over the axis. Based on those observations, we could predict that thermal conductivity may vary over the twisted angle and stacking sequence in the case of isotropic materials like $MoS_2$ and $MoSe_2$, owing to the change in Brillouin zone arrangement in the twisted configuration. To comprehend the proper mechanism of the heat transfer method in the t-BL system (AB, AB′, and intermediate twist angle), we have calculated the thermal conductivity of these systems by applying the optothermal Raman technique [8,27] by performing temperature and power dependent Raman measurements. This optothermal Raman approach has been proven to be a reliable and non-contact technique for determining the thermal conductivity of 2D materials [27]. This approach determines the exact location of a Raman active mode by concentrating a monochromatic laser beam onto the flakes. Thermal softening facilitates the redshift of Raman modes when the laser power or sample temperature increases. The shift rate can be employed for determining thermal conductivity in the t-BL system with the help of thermal modelling. Twisted bilayer (t-BL) $MoSe_2$ samples with different twist angles in the range of $0 \leq \theta \leq 60°$ were prepared by PMMA-based transfer technique. We are able to achieve t-BL of $MoSe_2$ that had various interlayer stacking angles by transferring $MoSe_2$ monolayers from one sapphire substrate as well as stacking them over another transferred $MoSe_2$ monolayer (1L) on $SiO_2$ [300 nm]/Si substrate. The equilateral triangle shape of ML $MoSe_2$ flakes helps us to measure the twist angles precisely (within an error limit of $< 1°$) using optical imaging. The optical microscope (OM) images and the height profiles of some selected t-BL flakes confirmed by AFM topography are shown in Figs. 1(a)-1(f). Their representative height profiles are shown in Fig. S1 (Supplemental Material).

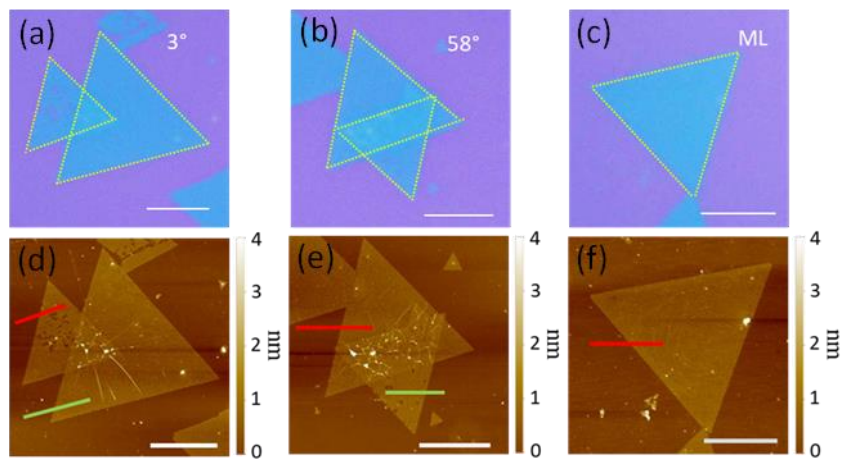

FIG. 1. [(a)-(c)] OM images of t-BL with stacking angles 3°, 58° and 1L $MoSe_2$ flakes (scale bar 5µm), respectively. (d-f) Corresponding AFM images of t-BL with stacking angles 3°,58° and 1L $MoSe_2$ flakes (scale bar 5µm), respectively.

In a bilayer TMDC system, there are five high-symmetry stacking patterns that can occur at angles 0° and 60°, depending on the crystal structure of the individual layers and the relative position of the layers. Two high-symmetry stacking patterns, 3R (often abbreviated as AB) and AA, are present at 0° and three stacking patterns that have an excellent symmetry, 2H (also known as AA′), AB′, and A′B are present at 60°. The stacking sequences such as AB and AB′ are known to be the most stable configuration, which is generally found in the triangular 2D TMDCs grown by the CVD method [41]. These stackings correspond to the mutual twist angles of 0° and 60°, respectively. Figs. 2(a)-2(b) shows the schematic of the stacking order in the top and side view. In artificial stacking, other intermediate twist angles are able to be found in addition to 0° and 60°. Hence for better comparison, the measurements have been performed on an intermediate twist angle *i.e.*31° along with the AB and AB′ stacking. In our measurements, the stacking angles have slightly deviated from the exact angle orientation, 0° and 60°. AA staking as2° and 3° whereas, for AB′, it is 57° and 58° respectively. This considerable deviation has occurred as the samples are made by artificial stacking, as discussed in the material and method section. Figs. 2(a)-2(b) shows the side and top view of high symmetric stacking AB which corresponds to the 0° stacking angle of bilayer MoSe2, where Mo is placed over Se, and the other Mo and Se are placed over the hexagonal centers. Similarly, Figs. 2(c)-2(d) shows the side and top view of high symmetric stacking AB′ which corresponds to the 60° stacking angle of bilayer MoSe2, where Mo is positioned over Mo and all of Se is positioned over the hexagonal centers.

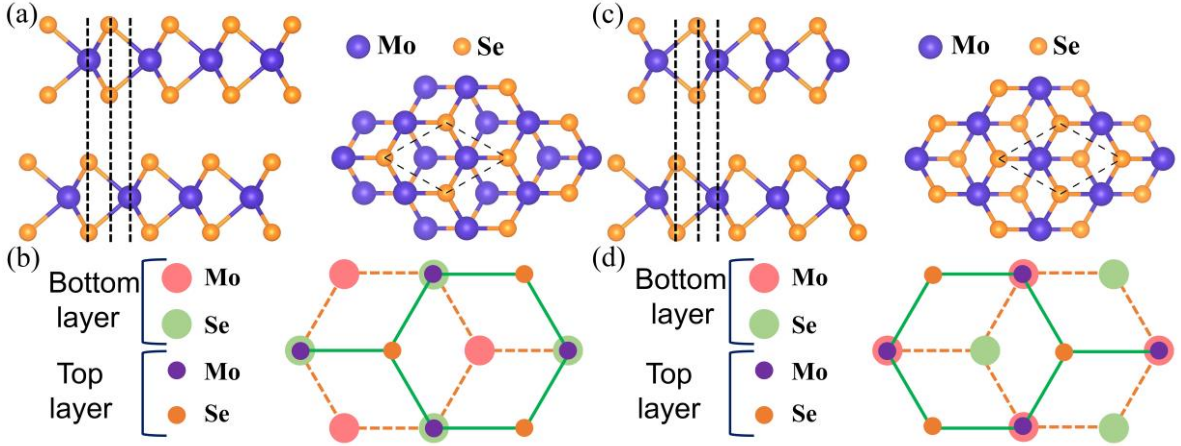

FIG. 2. [(a)-(b)] Side and top view of stacking AB and AB′ corresponds to 0° and 60° stacking angles of BL MoSe2, respectively. Purple (orange) solid sphere corresponds to Mo (Se) atoms. For each twist angle, a unit cell is shown, and its size varies strongly with the angle. [(c)-(d)] Top view of AB and AB′ stacking corresponds to 0° and 60° for BL MoSe2, respectively. Pink (green) solid circles correspond to Mo (Se) atoms at the bottom layer, and Purple (orange) circles correspond to Mo (Se) atoms at the top layer.

In order to characterize the t-BL MoSe2 and to confirm the stacking sequence, we have performed low frequency (LF) Raman and photoluminescence (PL) measurements. Usually, LF Raman measurement is considered a unique deterministic technique to visualize stacking orientation and twist angle in 2D systems [42]. Fig. 3(a) top and the bottom panel shows the LF Raman spectra for 58° and 3° twist angles as confirmed by the OM image [Figs. 1(a)-1(b)]. In the case of 58°, the shear mode (S) and breathing mode (B) appear at 18.9 cm$^{-1}$ and 28.9 cm$^{-1}$, respectively, whereas for 3° stacking, the S and B1 appear at 17.9 cm$^{-1}$ and 34.6 cm$^{-1}$. The separation between S and B is 11 cm$^{-1}$ and 10 cm$^{-1}$, which confirms the stacking sequence to be AB and AB′, as reported earlier [41]. Besides the mode separation, the relative intensity ratio S/B1 ~ 1/2 for 58° and S/B2 ~1/4 for 3° further confirms the stacking orientation as shown in the previous study for AB and AB′ stacking. B1 and B2 are specific types of breathing modes that can be observed in low-frequency Raman spectra. B1 corresponds to the symmetric breathing mode, while B2 corresponds to the antisymmetric breathing mode. In addition to the B2 mode, we have observed a significant contribution of the B1 mode for a 3° stacking angle. For AB′ (60°) or close to 60° (> 55°), this contribution of B2 mode usually disappears [41]. In the present scenario, 58° stacking does not show the signature of B2 as shown in Fig. 3(a) top panel. For more confirmation, we have performed LF Raman for one intermediate angle of 31° as shown in supplementary Fig. S3, where both B1 and B2 modes are present with a separation of 8 cm$^{-1}$. The disappearance of the S mode for the intermediate twist angle confirms the stacking sequence of AB and AB′ or close to 0° and 60° twist. Further, we have performed room temperature PL measurements for F2 and F1 flakes, as shown in Fig. 3(b) upper and lower panels, respectively. PL measurement is also performed on the exposed 1L part of the twisted sample for better comparison. In the case of 1L MoSe2, the main excitonic peak $AX^0$ is dominating around 789.5 nm [1.57 eV], whereas for the twisted bilayer case, $AX^0$ has been red shifted to 790.1 nm and 790 nm for 58° and 3° respectively. Besides, the intensity has been significantly reduced for the case of BL due to the indirect bandgap in nature. It can be noted that along with the A excitonic peak $AX^0$, trionic ($AX^-$) contribution is also observed for the twisted case. The energy separation between $AX^0$ and $AX^-$ is found to be 17 meV and 27 meV for the case of 58° and 3°, respectively. This energy separation of bilayer MoS2 (20 meV) shows a similar trend at room temperature, as reported earlier [43]. A significant change in the relative contribution of the exciton ($AX^0$) and trion ($AX^-$) has been observed for 3° and 58° stacking., which may be due to the change of stacking orientation or lattice mismatch. For intermediate angle 31°, $AX^0$ appears at ~ 790.9 nm as shown in supplementary Fig. S3, in which Trion peak ($AX^-$ at 803.2 nm) peak dominates with an energy separation of 23 meV.

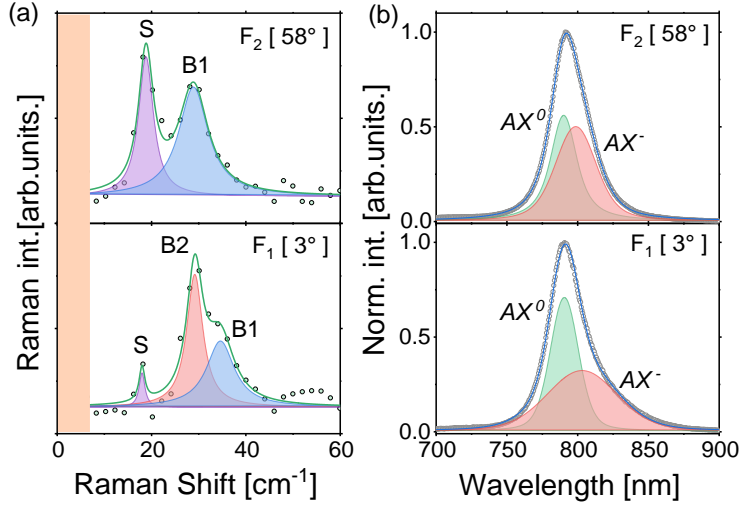

FIG. 3. (a) Stokes Raman spectra of the low-frequency shear modes (SM) and breathing modes (LBM) for two different twist angles, 3° and 58°, respectively. (b) Photoluminescence of the two different twist angles, 3° and 58°, respectively, which reflects changes in the exciton and trion contributions.

### A. Temperature dependent Raman spectra

A 632.8 nm incident laser was employed to conduct the temperature dependent Raman experiment for t-BL $MoSe_2$ and 1L $MoSe_2$. Fig. 4(a) shows the Raman spectra of nearly AB stacked (3°) t-BL, with the change in temperature from 296 to 476 K. The laser power was maintained below 15 μW to eliminate power dependent thermal heating and keep it constant during the course of the experiment. For the optimal peak position, the Lorentz function was fitted to each Raman spectra. The redshift of the Raman $A_{1g}$ peak is subsequently accompanied by a rise in temperature. Other twist angles of t-BL $MoSe_2$ also show a similar trend for Raman spectra, but the amount of redshift varies with different twist angles [Figs. 4(a)-4(b) and Figs. S3(a)-S3(c)]. Figs. 4(b)-4(c) shows the temperature dependent Raman spectra for t-BL (nearly AB′ stacked, 58°) and 1L $MoSe_2$, respectively.

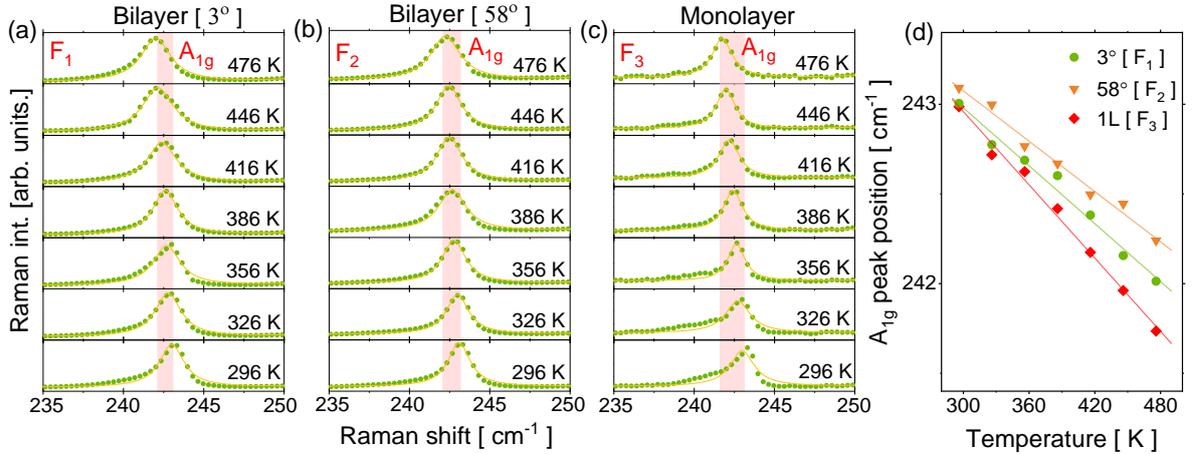

FIG. 4. [(a)-(c)] Raman spectra for t-BL system with stacking angles 3°, 58°and 1L $MoSe_2$ respectively, recorded in the temperature range 296 K to 476 K. (d) The temperature dependent Raman shift of $A_{1g}$ peak position for 3° (circle), 58° (triangle) and 1L $MoSe_2$ (square) respectively.

Similarly, such measurements have been carried out for other stacking angles 2°, 31°, 57° and another adjacent 1L MoSe2, shown in Fig. S3. Fig. 4(d) shows the variation of the $A_{1g}$ peak location in relation to temperature for t-BL systems along with 1L MoSe$_2$. The variation in the peak position of $A_{1g}$, with respect to temperature for other stacking angles 2°, 31°, 57° and 1L MoSe$_2$, are mentioned in Fig. S4. Being a function of lattice temperature, the modification that affects the Raman peak location ω [in cm$^{-1}$] exhibits a linear relationship [44], which is given by,

$$\omega(T) = \omega_0 + \chi_T T \quad (1)$$

where T represents temperature and the temperature coefficient of first-order is $\chi_T$. At absolute zero temperature, the vibration frequency corresponding to these phonon modes has been represented by $\omega_0$. The red shift that results from a temperature change (increase) can be attributed to thermally induced bond softening, as previously observed in graphene [8], which results in a change in Raman peak frequency. Hence, equation (1) becomes,

$$\Delta\omega = \omega(T_2) - \omega(T_1) = \chi_T(T_2 - T_1) = \chi_T \Delta T \quad (2)$$

where $\Delta\omega$ is the change in frequency, and $\Delta T$ is the change in temperature. The linear growth of phonon frequencies in t-BL and 1L MoSe$_2$ is driven by anharmonic lattice vibrations [45], which predominantly entail a contribution from the lattice's thermal expansion that resulted from the anharmonicity of the interatomic potential. The equilibrium sites of the atoms and, subsequently, the interatomic forces vary as the lattice expand or shrinks due to variations in temperature, which results in a change in the phonon energies [46].

### B. Power dependent Raman spectra

The significance of the connection between laser power and the $A_{1g}$ peak shift was investigated in the thermal conductivity study for both t-BL and 1L MoSe$_2$. Fig. 5(a) displays typical room temperature (RT) Raman spectra that were acquired using a 632.8 nm incident laser with powers ranging from 5μW to 63μW for t-BL MoSe$_2$ with stacking angle 3°. To determine the exact peak position, each Raman spectrum was fitted by employing the Lorentz function. Other stacking angles show a similar trend for Raman spectra, but the amount of red shift varies with different stacking angles. The local heating of the t-BL and 1L MoSe$_2$ causes the $A_{1g}$ Raman active mode to soften as laser power rises.

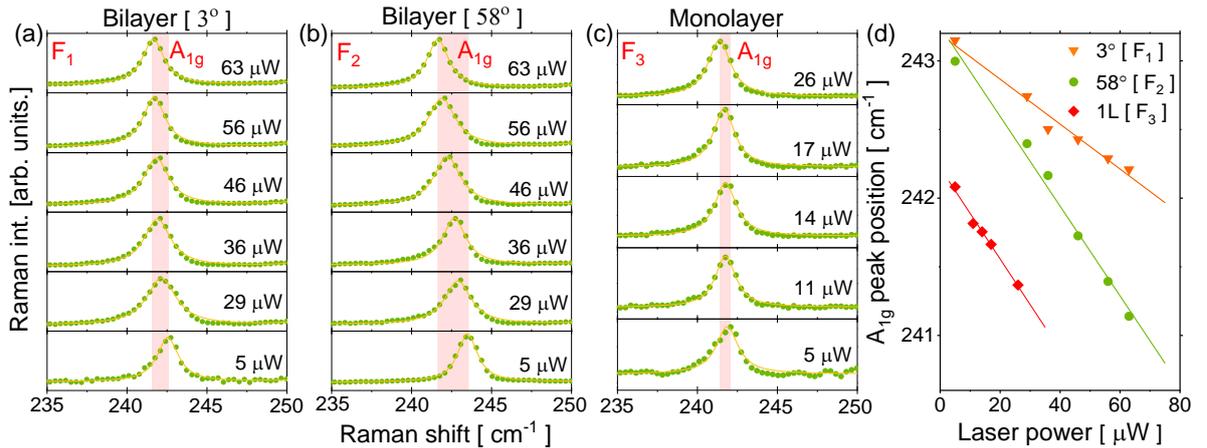

FIG. 5. [(a)-(c)] RT Raman spectra for the t-BL system with stacking angles 3°, 58° and 1L MoSe$_2$, respectively, recorded in the laser power ranging from 5μW to 63μW. (d) The power dependent $A_{1g}$ Raman peak shift for 3° (circle), 58° (triangle) and 1L MoSe$_2$ (square), respectively.

Throughout the course of this study, the laser power was consistently maintained below 0.1 mW (below the saturation limit) and kept constant throughout the measurement within the linear dependency range. Figs. 5(b)-5(c) shows the variation of the $A_{1g}$ peak location in relation to power for t-BL systems along with 1L MoSe$_2$. The variation in the peak position $A_{1g}$, with respect to power for other stacking angles 2°, 31°, 57° and another adjacent 1L MoSe$_2$, are mentioned in Fig. S6, where the A$_{1g}$ mode softens linearly with increasing power. The position of the peak as a function of power within the linear low-power range is depicted in Fig. 5(d) and Fig. S6, which is given by,

$$\Delta \omega = \omega(P_2) - \omega(P_1) = \chi_P (P_2 - P_1) = \chi_P \Delta P \quad (3)$$

Here, the power coefficient of first-order $\chi_P$ is defined by the variation in the phonon frequency $\omega$ with respect to incoming laser power P [$\delta\omega/\delta P$]. The thermal conductivity of this t-BL system was measured using a non-contact technique by taking advantage of MoSe$_2$'s phonon 'frequency's sensitive response to local laser heating. Despite not contributing yconsiderably to direct heat transfer through the flake, Raman active optical phonons demonstrate exceptional sensitivity to regional temperature fluctuation induced and regulated by external influences [47]. To determine the graphene's thermal conductivity, Balandin et al. [48] took into account the radial heat transfer and came up with the expression,

$$\kappa = (1/2\pi h)(\nabla P/\nabla T) \quad (4)$$

Where the change in the heating power $\nabla P$ is what causes the local temperature rise $\nabla T$ and the thickness of the material is represented by h. The thermal conductivity may be represented in the following manner by differentiating equation (1) with regard to power and replacing ($\nabla P/\nabla T$) into equation (4) which is provided by

$$\kappa = (1/2\pi h) \chi_T (\delta\omega/\delta P)^{-1} \quad (5)$$

where $\kappa$ is thermal conductivity, the temperature coefficient of first-order is given by $\chi_T$, and the phonon frequency fluctuation with incoming laser power is provided by ($\delta\omega/\delta P$). The sum of a material's 'lattice thermal conductivity ($\kappa_{latt}$) and 'electronic thermal conductivity' ($\kappa_e$) is equal to its total thermal conductivity ($\kappa$). The moment phonons, quantized lattice vibrations, are transported through a material, contributing to its thermal conductivity at the lattice level. However, the electrical component of thermal conductivity originates from the propagation of electrons and holes through the material. This contribution is typically dominant in metals and degenerate semiconductors, where the electrical conductivity is high. But for non-degenerate semiconductors, lattice thermal conductivity mostly contributes to total thermal conductivity ($\kappa \sim \kappa_{latt}$). For the purpose of determining the temperature and power coefficient of first-order for each twist angle, the power and temperature dependent Raman experiments were carried out, from which the thermal conductivity of t-BL MoSe$_2$ was calculated, which is shown in Table I.

TABLE I. Temperature coefficients of first-order, first-order power coefficients and room-temperature thermal conductivities of monolayer and t-BL MoSe$_2$

| Stacking angle (θ) | Temperature coefficient ($\chi_T$) (cm$^{-1}$/K) | Power coefficient ($\chi_P$) (cm$^{-1}$/μW) | Thermal conductivity [$\kappa$] W m$^{-1}$K$^{-1}$ |
|---|---|---|---|
| **Monolayer** | 0.00685 ± 0.00026 | 0.0339 ± 0.0021 | 38 ± 4 |
| **Bilayer [2°]** | 0.00543 ± 0.00026 | 0.0153 ± 0.0009 | 33 ± 3 |
| **Bilayer [3°]** | 0.00538 ± 0.00034 | 0.0163 ± 0.0013 | 31 ± 4 |
| **Bilayer [31°]** | 0.00504 ± 0.00026 | 0.0204 ± 0.0017 | 23 ± 3 |
| **Bilayer [57°]** | 0.00472 ± 0.00017 | 0.0311 ± 0.0033 | 14 ± 2 |
| **Bilayer [58°]** | 0.00468 ± 0.00025 | 0.0327 ± 0.0018 | 13 ± 1 |

It is notable from Table I that the thermal conductivity of t-BL MoSe$_2$ (TMDC) varies with different twist angles. Angle-dependent thermal conductivities of 2° and 3° (near AB stacking) have been calculated to be 33 ± 3 and 31 ± 4 W m$^{-1}$K$^{-1}$. Similarly, angle-dependent thermal conductivities of 57° and 58° (near AB′ stacking) have been calculated to be 14 ± 2 and 13 ± 1 W m$^{-1}$K$^{-1}$, respectively. For an intermediate angle (31°), the value is 23 ± 3 W m$^{-1}$K$^{-1}$ So, the thermal conductivity value for t-BL MoSe$_2$ is reducing from 33 to 13 W m$^{-1}$K$^{-1}$ with the change in twist angles. The moiré superlattices within this twisted configuration contribute extra phonon scattering, which decreases the mean free path of phonon and, consequently, the thermal conductivity. Apparently, when adjacent MoSe$_2$ layers are twisted, the number of atoms in the unit cell increases, which leads to an increment in the number of phonon vibration modes. This increment can reduce the phonon lifetime, as well as more phonons to scatter. In addition, the reduction in thermal conductivity found in t-BL MoSe$_2$ might be attributable to Brillouin zone changes [49]. These modifications are caused by in-plane rotation, which results in the formation of many folded phonon branches. The emergence of numerous folded phonon branches in t-BL MoSe$_2$ enhances both phonon Umklapp and normal scattering [50]. Umklapp phonon scattering occurs when phonons interact with each other and exchange momentum, resulting in the phonons being scattered in different directions. This scattering process can impede the heat flow and reduce the lattice thermal conductivity.

## IV. THEORYTICAL ANALYSIS

The thermal transport characteristics of BL MoSe$_2$ driven by stacking interactions were investigated by using first-principles density functional theory (DFT) computations. The objective was to theoretically study the effect of stacking on thermal transport. As the phononic properties are key factors for determining a material's thermal transport, the phonon band structure for the AB and AB′ stacked BL MoSe$_2$ has been computed and plotted in Figs. 6(a)-6(b), respectively.

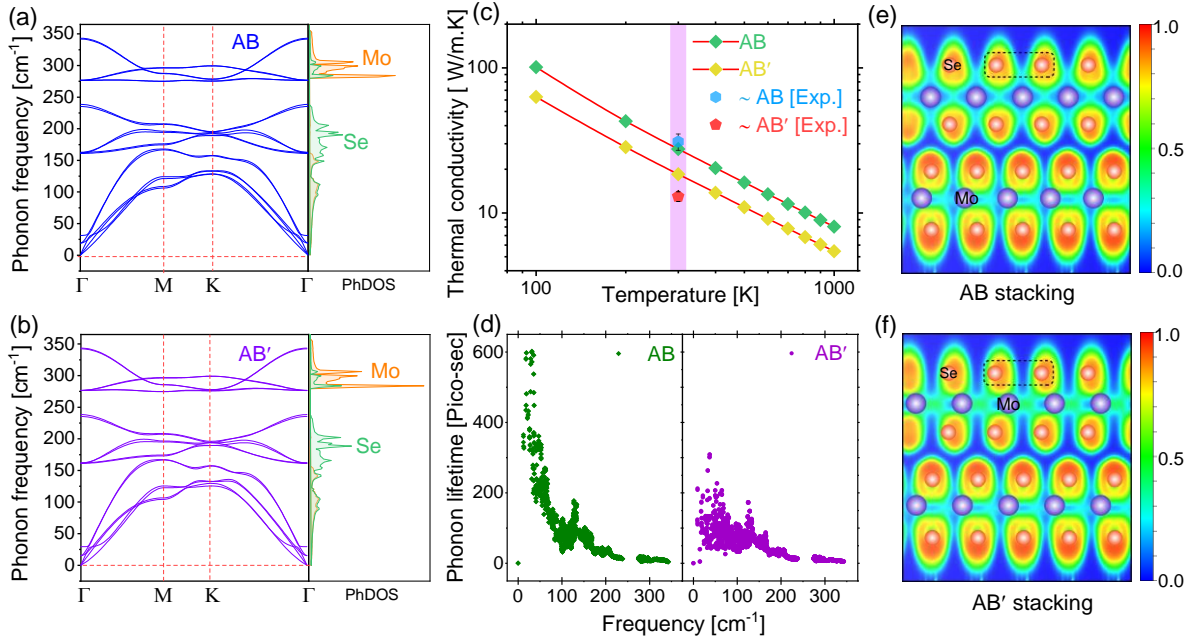

FIG. 6. Phonon dispersion (left panel) and PhDOS (right panel) for the high symmetric stacking (a) AB and (b) AB′ of BL MoSe$_2$ stacked at 0° and 60°, respectively. (c) Theoretically calculated thermal conductivity of BL MoSe$_2$ as a function of temperature for the stacking sequence of AB and AB′, respectively. The experimentally obtained thermal conductivities of the same stacking angles at RT are shown on highlighted (pink) region along the theoretical values. (d) Phonon lifetime of BL MoSe$_2$ for AB and AB′ stacking at RT. Corresponding projection of the electron localization function along (100) plane of (e) AB and (f) AB′ stacking order.

The phonon band structure study reveals the non-existence of imaginary phonon modes in the phonon dispersions, demonstrating the stability of dynamics in both stacking configurations. There exists a finite phononic band gap in both the stacking orders over the frequency of ~234 cm$^{-1}$, which plays an important role in determining the transport properties. Furthermore, the mass difference of Mo and Se in BL MoSe$_2$ indicates that the lower phonon frequency modes are provided by, the heavier cations Mo, and higher frequency modes by lighter anions Se, respectively. The phonon density of states (PhDOS) projected onto atoms was computed to evaluate the individual atoms' contributions towards phonon modes, as shown in the right panel of Figs. 6(a)-6(b) for AB and AB′ stacking orders, respectively. From the PhDOS of AB and AB′ stacked BL MoSe$_2$, it is observed that the low-frequency region in the range 0 to ~150 cm$^{-1}$ is contributed by both the Mo and Se atoms. Therefore, a hybridization of Mo and Se atoms is observed in this region, where all three acoustic modes, shear modes and breathing modes are present. In the intermediate region (frequency range ~150 to ~240 cm$^{-1}$), low-lying optical modes are present, including A$_{1g}$ Raman mode. The theoretically predicted position of the A$_{1g}$ Raman mode is 242 cm$^{-1}$, which is quite similar to the observed A$_{1g}$ mode peak position. In this intermediate region, the phonon vibration is solely contributed by the Se atom. In the high-frequency region, phonon modes (~275 to ~340 cm$^{-1}$) are caused by the vibrations of both the Mo and Se atoms. Similar to the low-frequency region, the hybridization of Mo and Se atoms in the lattice vibration is observed in this region. Finally, the lattice thermal conductivity ($\kappa_{latt}$) is calculated for AB and AB′ stacked BL MoSe$_2$ by computing the Boltzmann transport equation (BTE) for phonons. The values of $\kappa_{latt}$ for AB and AB′ stacking are 27.56 and 18.50 W m$^{-1}$K$^{-1}$, respectively, at room temperature (300K), which accord well with our experiments' results as plotted in Fig. 6(c). Therefore, a large reduction of $\kappa_{latt}$ is observed by changing the stacking order from AB to AB′ in BL MoSe$_2$, as observed from our experimental as well as theoretical calculations. The temperature effect on the thermal transport of AB and AB′ stacked BL MoSe$_2$, the $\kappa_{latt}$, is calculated at various temperatures in the 100 to 1000 K range. For AB stacking, the value of $\kappa_{latt}$ decreases from 101.02 to 8.04 W m$^{-1}$K$^{-1}$ when temperature varies from 100 to 1000 K, nearing the theoretical minimum value, with a predicted 1/T dependency [51]. For AB′ stacking, this value decreases from 63.04 to 5.45 W m$^{-1}$K$^{-1}$ when the temperature changes from 100 to 1000 K. The variation of $\kappa_{latt}$ with temperature is demonstrated in Fig. S7 (a) and (b) for AB and AB′ stacked BL MoSe$_2$, respectively. With the increase in temperature, the value of $\kappa_{latt}$ decreases very fast in the low-temperature range and almost saturates beyond a certain temperature (~500 K). This is due to the rapid increase of thermal scattering with temperature, which saturates after a certain higher value of temperature. The log-log plot of $\kappa_{latt}$ with respect to temperature is shown in Fig. 6(c), which reveals the similarity of thermal transport in both the stacking orders and indicates that both the stacking order possesses identical variation of $\kappa_{latt}$ with temperature by a constant shift of $\Delta\kappa_{latt}$ in the log scale. Although the different stacking order makes the difference in $\kappa_{latt}$ value, it has a similar effect of temperature on $\kappa_{latt}$ as well as the thermal scattering. In order to elucidate the fundamental root cause of the stacking-dependent lattice thermal conductivity ($\kappa_{latt}$) of bilayer MoSe$_2$, we have considered the kinetic theory of lattice thermal conductivity:

$$\kappa_{latt} = \frac{1}{3}C v_g^2 \tau \qquad (6)$$

where $C$ demonstrates the lattice-specific heat capacity, $v_g$ denotes the average group velocity of phonons, and $\tau$ represents the average relaxation time of phonons. As the stacking order in the multi-layer and heterostructure determine the intralayer and interlayer interactions between the constituent atoms [52,53], it has a significant effect on the phonon scattering. Therefore, both the lifetime ($\tau$) and group velocity ($v_g$) of phonon, two fundamental physical parameters, might be affected by the change in stacking order. To further understand the influence of stacking order on thermal conductivity, the electron localization function (ELF) and phonon lifetime were determined.

### A. Phonon lifetime

The phonon lifetime for AB and AB′ stacked MoSe$_2$ at 300 K is shown in Fig. 6(d) left and right panels, respectively. The phonon lifetime represents the total time for heat carrying by phonons before the scattering. Therefore, $\kappa_{latt}$ has a direct proportionality relation with phonon lifetime ($\tau$) as indicated in equation [6]. Therefore, the higher $\kappa_{latt}$ results from the longer phonon lifetime [54,55]. Phonon lifetime for AB stacking is found in the range of 0 to ~600 ps, whereas, for AB′ stacking, it is in the range of 0 to ~300 ps. Therefore, the maximum limit of the phonon lifetime is reduced significantly when stacking order changes from AB to AB′ in BL MoSe$_2$. Hence, the average heat carrying by phonons in the AB stacking is more than AB′ stacking, leading to higher $\kappa_{latt}$. This is also applicable to other temperatures of the system. As a result of this at any temperature in the range (100 to 1000 K), the value of $\kappa_{latt}$ is more in AB stacking than AB′ stacking.

### B. Electron localization function (ELF)

The electron localization function (ELF) [56,57] was studied in order to better comprehend the bonding in the bilayer system. ELF estimates the likelihood of finding an electron close to the reference electron. The limitations for entirely localized and extremely delocalized electrons are represented by the ELF's values of 1 and 0, respectively. The ELF projected along [100] for AB and AB′ stacking has been shown in Figs. 6(e)-6(f), respectively. In both stackings, the bottom layer is fixed, while the top layer is displaced and rotated with respect to the bottom layer depending upon the stacking order. Therefore, we can find the identical ELF of the lower layer in both the stacking order, whereas there is finite deference between the ELF of the top layer of AB and AB′ stacking, as shown in Figs. 6(e)-6(f). This difference creates differences in the thermal transport properties of the AB and AB′ stacking. For the $\kappa_{latt}$, the low frequency and medium frequency range of lattice vibration plays a major role. Our experiment results also suggest the contribution of A$_{1g}$ Raman mode on the $\kappa_{latt}$, which lies in the medium frequency range. In this frequency range, mainly the vibration is due to the Se atoms, as shown in the right panel of Figs. 6(a)-6(b) for AB and AB′ stacking, respectively. Therefore, the ELF between consecutive Se atoms is a major concern for the thermal transport of the lattice vibration. From Fig. 6(e), it is observed that in between the Se atoms, there exists a finite electron localization in AB stacking order (dotted black box region). On the contrary, the electron localization in AB′ stacking order is significantly weak, as shown in Fig. 6(f). The high electron localization facilitates the $\kappa_{latt}$ in the system [55]. The relatively stronger electron localization increases the phonon group velocity and assists to increase $\kappa_{latt}$ with the phonon lifetime for AB stacking r than AB′ stacking. Along with AB and AB′ stacking, Fig. S8 also depicts the electron localization for the intermediate stacking angle. From the electron localization for the intermediate stacking angle, it can be seen that the strength of the electron localization between Se atoms is stronger than AB′ but weaker than AB stacking. This is the reason for higher and lower values of thermal conductivity in intermediate stacking angle than AB′ and AB stacking, respectively.

### V. CONCLUSION

We have synthesized t-BL MoSe$_2$ with different twist angles using APCVD followed by the wet transfer method. The optothermal Raman approach was implemented to calculate the thermal conductivity of 1L and t-BL systems at ambient temperature. The thermal conductivity varies as a function of different twist angles has been addressed for the first time. The observed thermal conductivity values are found to be 13 ± 1, 23 ± 3 and, 30 ± 4 W m$^{-1}$K$^{-1}$ for twist angle [θ] = 58°, 31°and 3° respectively. The thermal conductivity of 1L (38 ± 4 W m$^{-1}$K$^{-1}$) is found to be larger than the bilayer due to the decrease in phonon scattering at the interface between layers. Theoretically, the estimated thermal conductivity of bilayer MoSe$_2$ for AB and AB′ stacking orders using the phonon Boltzmann transport equation (BTE) strongly supports the experimental results. Furthermore, the theoretical DFPT calculations and ELF analysis explore the fundamental origin of twist angle-dependent thermal conductivity in t-BL MoSe$_2$. The change in twist angle in t-BL MoSe$_2$ exhibits the twist angle-driven thermal conductivity, which may be attributed to changes in both the group velocity and the lifetime of phonons. The findings show more precise thermal transport measurements in the t-BL (homo-bilayer 2D) system, a significant paradigm for device modelling and other applications.

ACKNOWLEDGMENTS

P.K.N. acknowledges the financial support from the Ministry of Human Resource Development (MHRD), Government of India (GOI) via STARS grant [STARS/APR2019/148] and Department of Science and Technology, Government of India (DST-GoI), with sanction Order No. SB/S2/RJN-043/2017 under Ramanujan Fellowship. P. K. N. also acknowledges the support from the Institute of Eminence scheme at IIT-Madras, through the 2D Materials Research and Innovation Group and Micro Nano Bio Fluidics group.

# Supplemental Material

# Probing angle dependent thermal conductivity in twisted bilayer MoSe$_2$

Manab Mandal[1,2], Nikhilesh Maity[3], Prahalad Kanti Barman[2], Ashutosh Srivastava[3], Abhishek K. Singh[3], Pramoda K. Nayak[2,4*], Kanikrishnan Sethupathi[1,2*]

[1] Low temperature physics lab, Department of Physics, Indian Institute of Technology Madras, Chennai 600 036, India
[2] Department of Physics, Indian Institute of Technology Madras, Chennai 600 036, India
[3] Materials Research Centre, Indian Institute of Science, Bangalore-560012, India
[4] Centre for Nano and Material Sciences, Jain (Deemed-to-be University), Jain Global Campus, Kanakapura, Bangalore 562112, India

*Corresponding authors: ksethu@iitm.ac.in, pramoda.nayak@jainuniversity.ac.in

List of Supplemental Figures

FIG. S1: Optical images and AFM topography of $F_4$, $F_5$, $F_6$ and $F_7$ flakes.

FIG. S1A: AFM height profiles of $F_1$, $F_2$, $F_3$ flakes.

FIG. S1B: AFM height profiles of $F_4$, $F_5$, $F_6$ and $F_7$ flakes.

FIG. S2: Low-frequency Raman spectra and Photoluminescence spectra of $F_5$ flake.

FIG. S3: Temperature dependent Raman spectra of $F_4$, $F_5$, $F_6$ and $F_7$ flakes.

FIG. S4: Position of $A_{1g}$ Raman peak vs. temperature for $F_4$, $F_5$, $F_6$ and $F_7$ flakes.

FIG. S5: Power dependent Raman spectra of $F_4$, $F_5$, $F_6$ and $F_7$ flakes.

FIG. S6: Position of $A_{1g}$ Raman peak vs. laser power for $F_4$, $F_5$, $F_6$ and $F_7$ flakes.

FIG. S7: Theoretically calculated thermal conductivity for the stacking sequence of AB and AB′

FIG. S8: Electron localization function AB′, 21.8º and, AB stacking order.

FIG. S9: Comparative study of thermal conductivity.

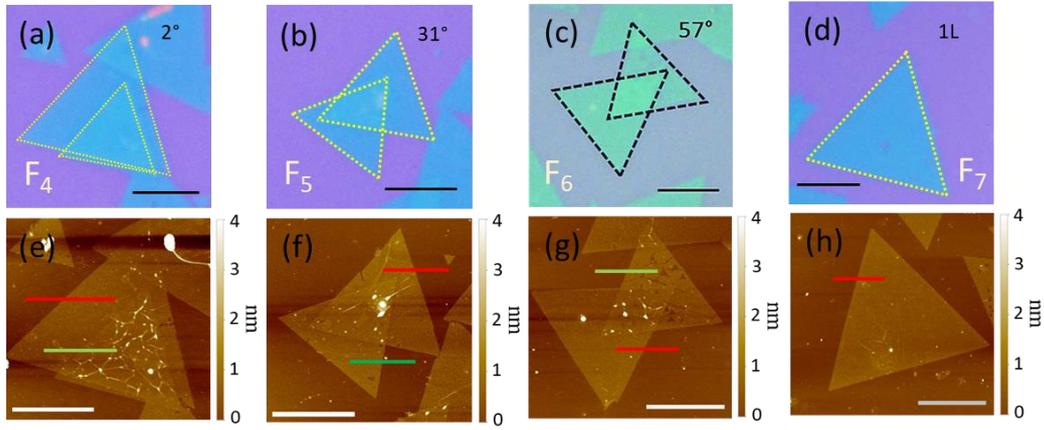

FIG. S1. [(a)-(d)] Optical reflection images of twisted bilayer with stacking angles 2°, 31°, 57° and monolayer $MoSe_2$ flakes, respectively. [d-f] Atomic force microscopy images of twisted bilayer with stacking angles 2°, 31°, 57° and monolayer $MoSe_2$ flakes respectively. Scale bar is 5 µm for all the images.

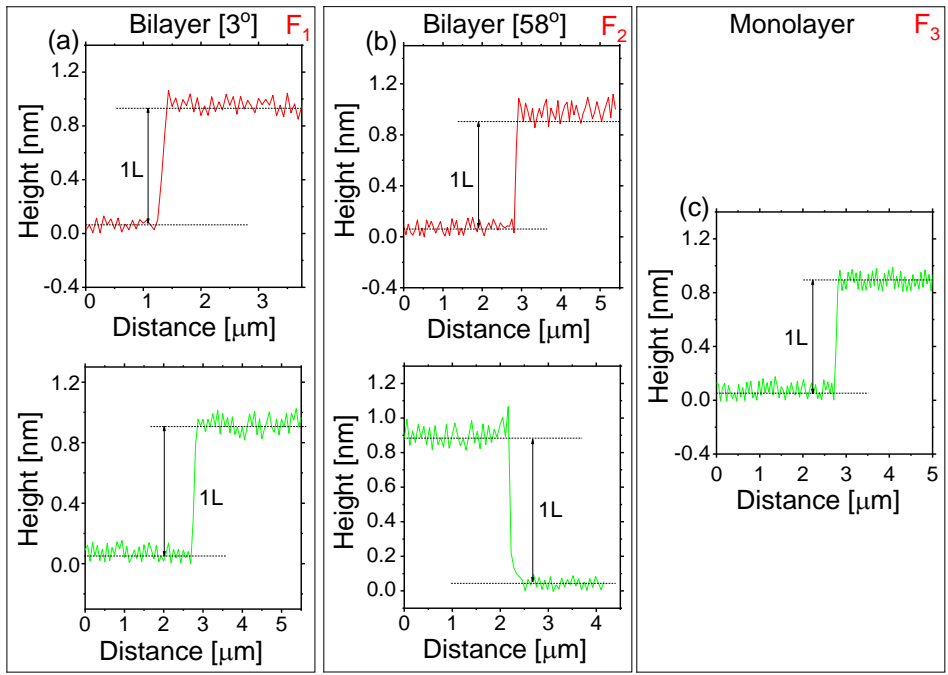

FIG. S1A. [(a)-(c)] AFM height profile of a triangular flakes, showing a step height of ~ 0.83 nm corresponds to twisted bilayer (green ~ bottom layer thickness profile and Red ~ top layer thickness profile) with stacking angles 3°, 58° and monolayer $MoSe_2$.

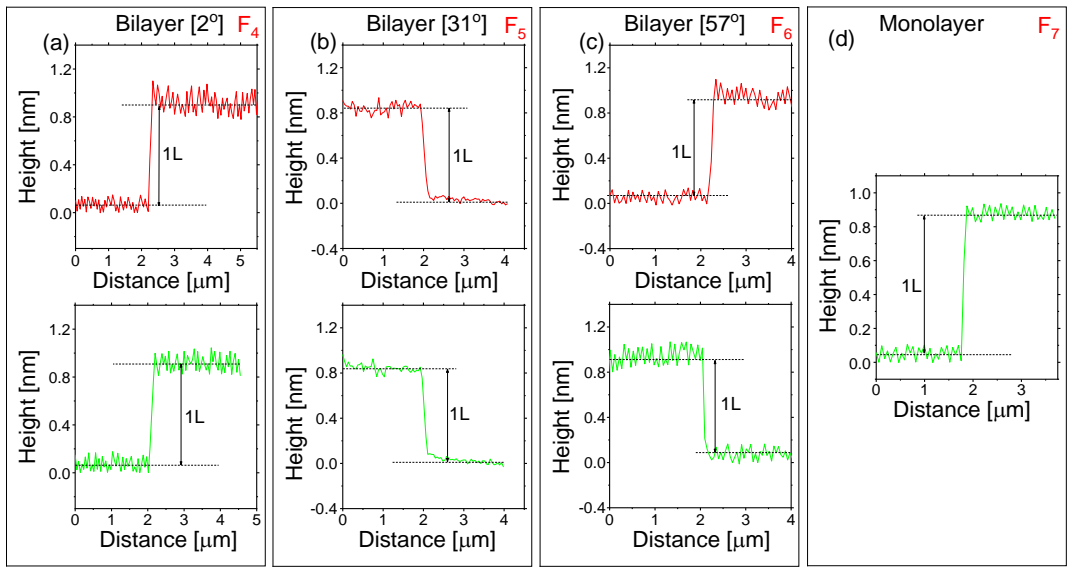

FIG. S1B. [(a)-(d)] AFM height profile of a triangular flakes, showing a step height of ~ 0.83 nm corresponds to twisted bilayer (green ~ bottom layer thickness profile and Red ~ top layer thickness profile) with stacking angles 2°, 31°, 57° and monolayer $MoSe_2$.

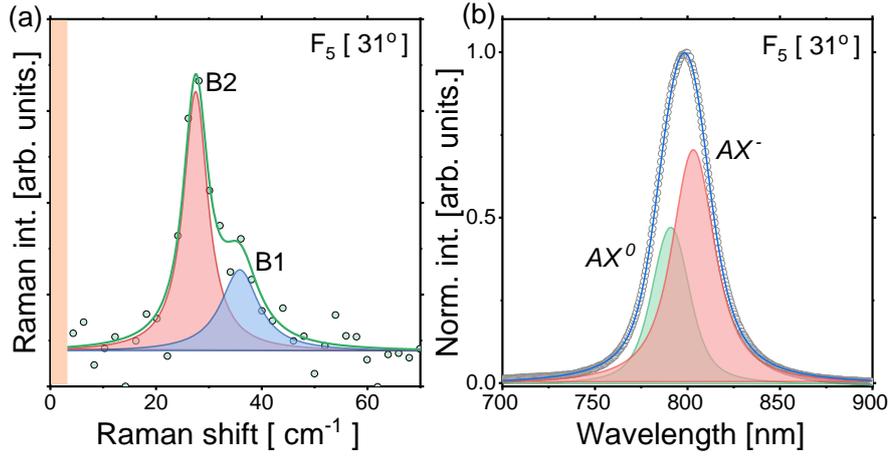

FIG. S2. (a) Stokes Raman spectra of the low-frequency shear (S) and breathing modes (B1 & B2) for twisted bilayer $MoSe_2$ flakes stacked at 31°. (b) Photoluminescence spectra of the same twisted bilayer $MoSe_2$ flakes stacked at 31°. The contribution of neutral exciton ($AX^0$) and trion ($AX^-$) are marked.

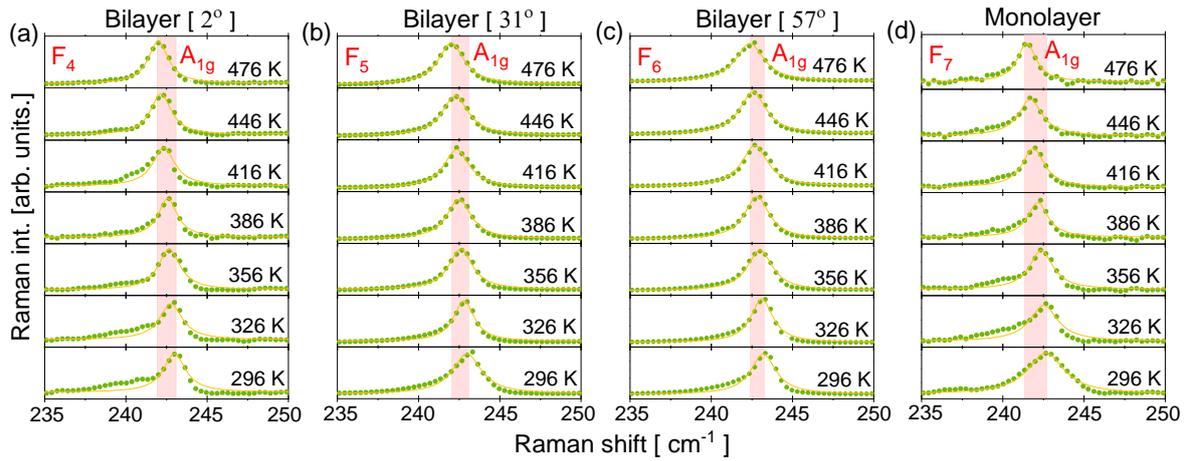

FIG. S3. [(a)-(d)] Raman spectra for t-BL $MoSe_2$ with stacking angles 2°, 31°, 57° and monolayer $MoSe_2$ respectively, recorded at different temperatures in the range 296 K to 476 K. The symbols represent experimental data and the solid lines are the fitting data.

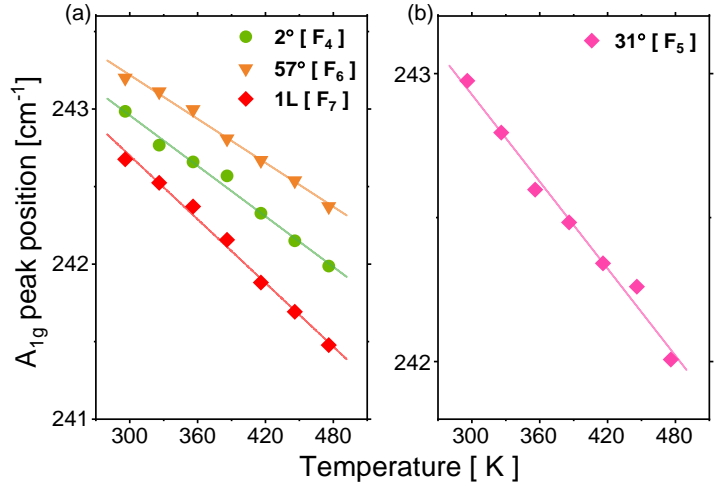

FIG. S4. (a) Position of $A_{1g}$ Raman peak vs. temperature of t-BL MoSe$_2$ with stacking angles 2°, 57° and monolayer MoSe$_2$. (b) t-BL MoSe$_2$ system with stacking angle 31°. The symbols represent experimental data and the solid lines are the fitting data.

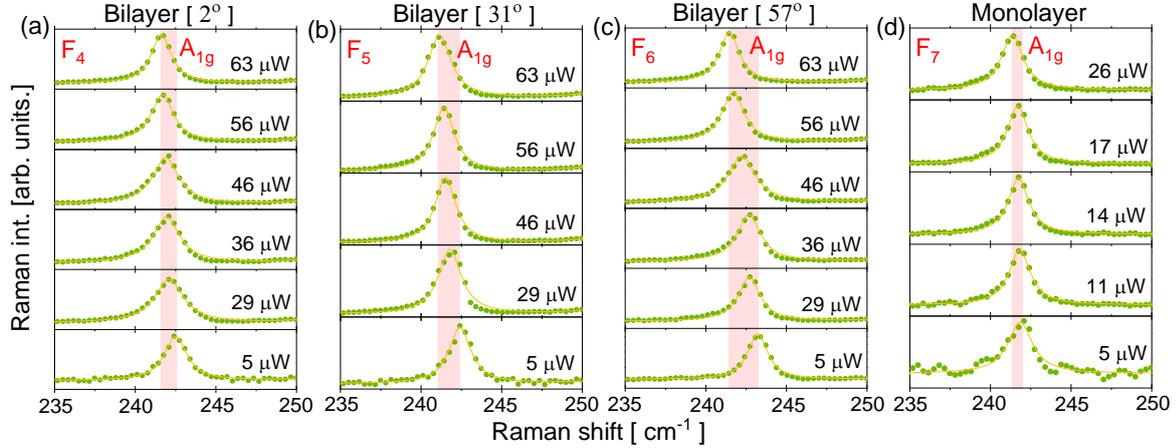

FIG. S5. [(a)-(d)] Raman spectra for t-BL MoSe$_2$ with stacking angles 2°, 31°, 57° and monolayer MoSe$_2$ respectively, recorded at different powers in the range 5 μW to 63 μW. The symbols represent experimental data and the solid lines are the fitting data.

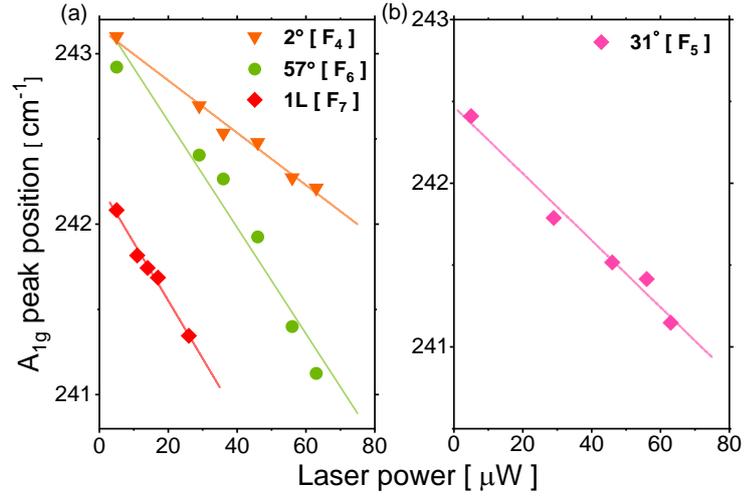

FIG. S6. (a) The position of $A_{1g}$ Raman peak vs. laser power for t-BL MoSe$_2$ with stacking angles 2°, 57° and monolayer MoSe$_2$ and (b) t-BL MoSe$_2$ system with stacking angle 31°. The symbols represent experimental data and the solid lines are the fitting data.

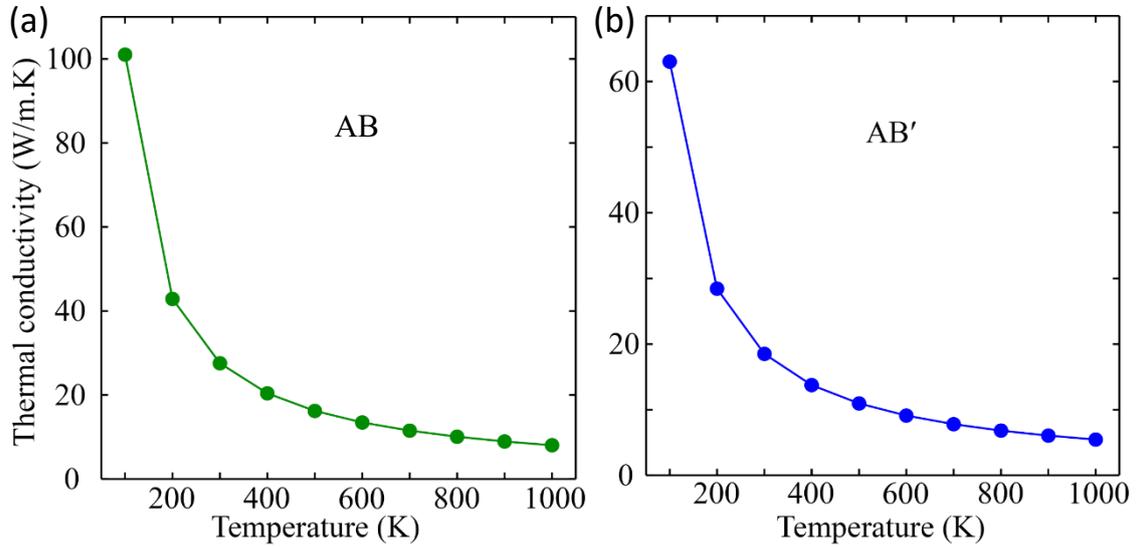

FIG. S7. Theoretically calculated thermal conductivity of BL MoSe$_2$ as a function of temperature for the stacking sequence of AB (a) and AB′ (b) respectively.

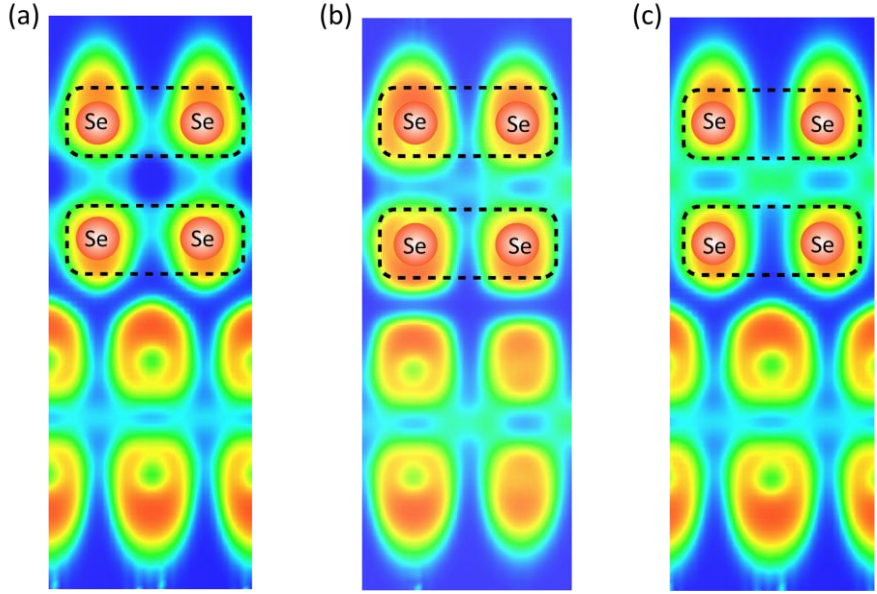

FIG. S8. Corresponding projection of the electron localization function along (100) plane of (a) AB′, (b) 21.8º and, (c) AB stacking order.

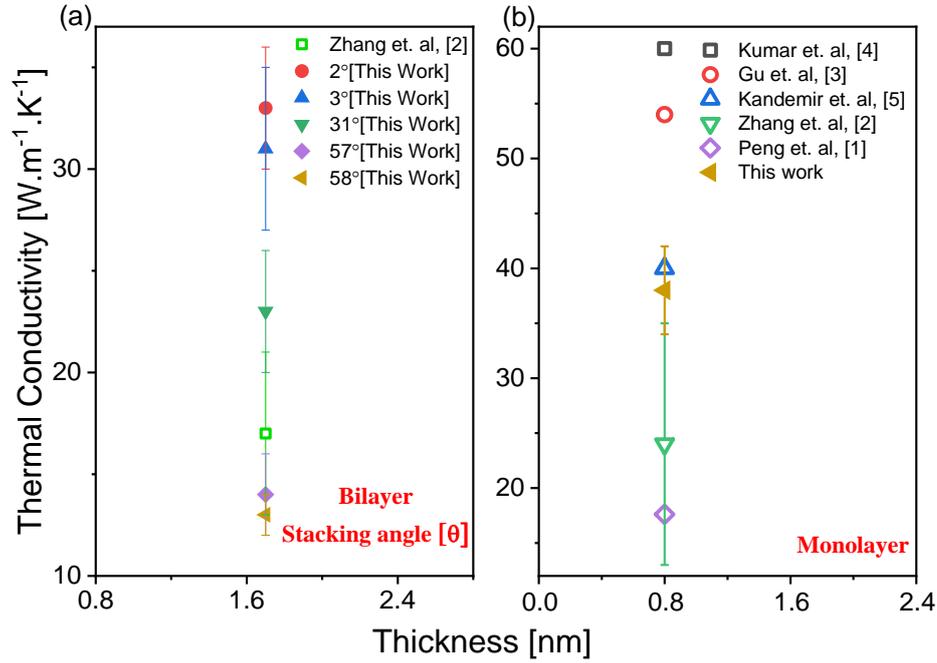

FIG. S9. [(a)-(b)] Theory and Experiment vs literature. Thermal conductivity of bilayer and monolayer $MoSe_2$ at 300 K obtained in this work (Solid symbol) and reported in the literature (Hollow symbol) by Peng et al., [1] Zhang et al., [2] Gu et al., [3] Kumar et al., [4] Kandemir et al., [5] respectively.